\documentclass{aa}
\newcommand{\bit}{\begin{itemize}}
\newcommand{\eit}{\end{itemize}}
\newcommand{\ben}{\begin{enumerate}}
\newcommand{\een}{\end{enumerate}}
\newcommand{\bde}{\begin{description}}
\newcommand{\ede}{\end{description}}

\usepackage{graphicx}
\usepackage{lscape}

\begin{document}
\title{Internal kinematics of modelled interacting disc galaxies}
\author{T. Kronberger$^{1,2}$,
        W. Kapferer$^1$,
        S. Schindler$^1$,
        A. B\"ohm$^{2,3}$,
        E. Kutdemir$^2$, and
        B. L. Ziegler$^2$}

\institute{ $^1$Institut f\"ur Astro- und Teilchenphysik,
            Universit\"at Innsbruck,
            Technikerstr. 25,
            A-6020 Innsbruck, Austria \\
            $^2$Institut f\"ur Astrophysik,
            Universit\"at G\"ottingen,
            Friedrich-Hund-Platz 1,
            D-37077 G\"ottingen, Germany \\
            $^3$Astrophysikalisches Institut Potsdam,
            An der Sternwarte 16,
            D-14482 Potsdam, Germany}

\offprints{T. Kronberger, \email{Thomas.Kronberger@uibk.ac.at}}

\date{-/-}

\abstract{We present an investigation of galaxy-galaxy
interactions and their effects on the velocity fields of disc
galaxies in combined N-body/hydrodynamic simulations, which
include cooling, star formation with feedback, and galactic winds.
Rotation curves (RCs) of the gas are extracted from these
simulations in a way that follows the procedure applied to
observations of distant, small, and faint galaxies as closely as
possible. We show that galaxy-galaxy mergers and fly-bys disturb
the velocity fields significantly and hence the RCs of the
interacting galaxies, leading to asymmetries and distortions in
the RCs. Typical features of disturbed kinematics are
significantly rising or falling profiles in the direction of the
companion galaxy and pronounced bumps in the RCs. In addition,
tidal tails can leave strong imprints on the rotation curve. All
these features are observable for intermediate redshift galaxies,
on which we focus our investigations. We use a quantitative
measure for the asymmetry of rotation curves to show that the
appearance of these distortions strongly depends on the viewing
angle. We also find in this way that the velocity fields settle
back into relatively undisturbed equilibrium states after unequal
mass mergers and fly-bys. About 1 Gyr after the first encounter,
the RCs show no severe distortions anymore. These results are
consistent with previous theoretical and observational studies. As
an illustration of our results, we compare our simulated velocity
fields and direct images with rotation curves from VLT/FORS
spectroscopy and ACS images of a cluster at z=0.53 and find
remarkable similarities.

}

\authorrunning {T. Kronberger et al.}
\titlerunning {Internal kinematics of modelled interacting disc galaxies}

\maketitle
%

\section{Introduction}
The evolution of galaxies within the dense environment of a galaxy
cluster is strongly influenced by various interaction phenomena.
These environmental effects are superposed on the hierarchical
growth predicted by cold dark matter (CDM) theories. Tidal
interactions, for instance, have a higher probability in galaxy
clusters and groups, due to the higher number density of galaxies.
Going to higher redshifts, the merger frequency increases
(Dressler et al. 1994), as galaxy clusters are assumed to assemble
rather late in hierarchical CDM models. The relative velocities
during encounters in galaxy clusters are higher than in the field
and therefore reduce the probability for mergers.

In a recent paper we studied the influence of tidal interactions
and mergers on the star formation rates of the involved galaxies
using numerical simulations (Kapferer et al. 2005). However, not
only is the star formation affected by galaxy-galaxy mergers, but
also the internal kinematics of the gas. Most of the
irregularities in the velocity field reflect disturbances in the
gravitational potential of the galaxy. It is expected that
kinematic disturbances fade within a few rotation cycles ($\le$ 1
Gyr) (e.g. Dale et al. 2001). Therefore, they can be used to trace
the recent interaction history of a galaxy and give possible clues
to the type of the respective interaction. In addition,
identifying disturbances in the RCs is important for Tully-Fisher
studies in order to accurately derive the maximum rotation
velocity V$_{\rm max}$.

Observations by Rubin, Waterman, \& Kenney (1999) have shown, for
example, that rotation curves from several Virgo cluster galaxies
exhibit clear kinematic disturbances. However, not all
peculiarities are severe. Some rotation curves show only small
asymmetries and others have no distortions at all. The authors did
not find any strong correlations with Hubble type, local galaxy
density, or HI deficiency. Many other studies have also relied on
1D information about the internal kinematics and used RCs to trace
interactions in galaxy groups and clusters (e.g. Swaters et al.
1999; Dale et al. 2001; Conselice \& Gallagher 1999; Mendes de
Olivera et al. 2003). During the past few years it has become
feasible to observe the full 2D velocity field of \textit{local}
galaxies. For example Chemin et al. (2006) investigated the
H$\alpha$ line of 30 spiral and irregular galaxies in the Virgo
cluster using Fabry-Perot interferometry (FPI). Their images show
different signs of interactions in several objects. Likewise, the
GHASP survey observed local field galaxies (Garrido et al.
\cite{Garrido}) with FPI.

Almost all studies of distant, faint, and small galaxies are still
based on slit spectroscopy (Vogt 2001, B\"ohm et al. 2004).
Focussing on the environment of clusters, Ziegler et al. (2003)
observed spirals in three intermediate redshift clusters
(z=0.3-0.6), using multi-object spectroscopy with VLT/FORS. In the
same manner, Bamford et al. (2005) investigated five distant
clusters. Using Subaru/FOCAS, Nakamura et al. (2006) examined four
distant clusters. Finally, Metevier et al. (2006) studied a
cluster at z=0.4 using Keck/LRIS. All these studies analysed the
luminosity evolution of the galaxies via the Tully-Fisher
relation. While Ziegler et al. (2003) and Nakamura et al. (2006)
do not find significant differences between galaxies in clusters
and in the field, Bamford et al. (2005) claim that cluster
galaxies are on average brighter than their field counterparts.
Part of this discrepancy may be attributed to the way distortions
and irregularities in the RCs are accounted for.

Therefore, to systematically explore the influence of interactions
on the velocity fields and hence the RCs, we present here a
comprehensive analysis using numerical simulations. We present the
results by closely mimicking the observations cited above. To this
purpose we extract RCs that can be directly compared to the
observed ones. Our N-body/SPH simulations have already been
successfully applied to model RCs of isolated spiral galaxies
(Kapferer et al., 2006). In that work we examined, for example,
how observational constraints affect the quality of RCs. Here we
go one step further to investigate the kinematics of tidally
interacting and merging systems. Depending on the interaction
geometry, galaxy mergers and fly-bys will cause distortions of
different strengths in the velocity field of galaxies.

The main questions we will answer are what kind of distortions in
the RCs can arise from tidal interactions? Are these peculiarities
observable at intermediate and high redshifts? Are the features
permanently visible or do they vanish within certain timescales?
Some of the questions that we investigate in the present work were
also addressed by Barton et al. (\cite{Barton}) and Heyl et al.
(1996), who found qualitatively similar results to ours. However,
as both groups used purely N-body models, they were not able to
extract RCs as close to the observational procedure as we do (see
Sect. 2.1). This is especially crucial in the treatment of high
redshift galaxies, which we focus on in the present study. A very
interesting work connecting observations and simulations was
presented by Bournaud et al. (2004), who investigated tidal tails
in interacting galaxies, also with respect to the signatures they
leave in the position-velocity diagram of the galaxy. They found
that, depending on the viewing angle, the sign of the velocity
gradient can change in the presence of tidal tails.

This paper is organised as follows. In Sect. 2 and 3, we describe
the simulations and the way we extract realistic rotation curves
from the numerical data. Sect. 4 explains the interaction
geometries used for the present work. In Sect. 5 the results for
different interaction scenarios and their dependence on the
viewing angle are presented, followed by a short discussion of
some of the results' implications for observational properties of
cluster galaxies in Sect. 6. We conclude with a summary of the
main conclusions in Sect. 7.

\section{Simulations}
In this work we use some of the interacting systems presented in
Kapferer et al. (2005). The simulations were carried out with the
N-body/SPH code GADGET-2 developed by V. Springel (see Springel
2005 for details). In this code the gas of the galaxies is treated
hydrodynamically, including cooling, star formation, stellar
feedback, and galactic winds (Springel \& Hernquist, 2003). The
collisionless dynamics of the dark matter and the stellar
component is calculated using an N-body technique. The combined
N-body/SPH simulation then calculates 5 Gyr of evolution. For
every time step, we know the velocity of each particle and can
extract realistic rotation curves. Other authors have used purely
N-body techniques to investigate the kinematics of the modelled
galaxies (e.g. Barton et al. \cite{Barton}; Heyl et al. 1996).
Although the overall kinematics of the disc is dominated by the
darkmatter halo, disturbances of the gas have to be treated
hydrodynamically, especially with respect to survival timescales
of the distortions. The dynamics of the gas during an interaction
is significantly different from the dynamics of the collisionless
stellar component. Thus, a purely collisionless approach is
insufficient for describing the kinematics of the gas in detail.
Note that also the star formation rate of the interacting system
is usually higher than that of the isolated system (Kapferer et
al., 2005). Therefore, stronger galactic winds can occur, which
will, in general, decrease the rotational velocities (see paper I
for a discussion of this effect). Nevertheless, the results
obtained by the above-cited groups are very interesting and in
several points similar to ours, so we will use them frequently as
reference in subsequent sections. Iono et al. (2004) also used an
N-body/SPH code to investigate position velocity diagrams of
modelled galaxies, but focussed on signatures of radial gas flows.

In Kapferer et al. (2006, hereafter paper I) we have shown that
our models successfully reproduce rotation curves of isolated
spiral galaxies. For completeness we repeat the initial conditions
of our model galaxies in Table \ref{galprop}. These initial
conditions were built according to Springel et al. (2005), which
is based on the work of Mo et al. (1998). The two model galaxies
were chosen such that they represent a Milky Way type and a small
spiral galaxy, where the mass ratio of the two galaxies is 8:1.

\begin{table}
\caption[]{Properties of the initial conditions of the model
galaxies.}
\begin{tabular}{l l l l }
\hline \hline Properties & Galaxy A & Galaxy B  \cr \hline
circular velocity $V_{\rm{c}}$$^{1}$& 160 & 80 \cr disc mass
fraction$^{2}$ & 0.05 & 0.05  \cr gas content in the disc$^{3}$ &
0.25 & 0.25  \cr disc thickness$^{4}$ & 0.02 & 0.02  \cr  total
mass [$M_{\sun}$] & 1.33x10$^{12}$ $h^{-1}$ & 1.67x10$^{11}$
$h^{-1}$ \cr disc scale length [kpc] & 4.51$h^{-1}$ & 2.25$h^{-1}$
\cr \hline
\end{tabular}
\label{galprop} $^{1}$... circular velocity at $\rm{r}_{200}$ in
km/s\newline $^{2}$... fraction of disc particles (stars/gas) in
units of halo mass \newline $^{3}$... relative content of gas in
the disc
\newline $^{4}$... thickness of the disc in units of radial scale
length
\newline
\end{table}

Throughout the paper we adopt the standard $\Lambda$CDM cosmology
with $\Omega_\Lambda=0.7$, $\Omega_m=0.3$, and h=0.7.

\subsection{Advanced rotation curve extraction}
\begin{figure}
\begin{center}
{\includegraphics[width=\columnwidth]{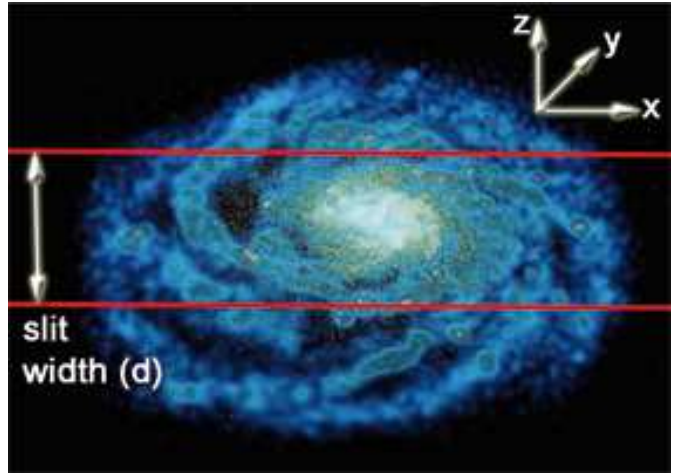}} \caption{Model
galaxy A (gas (blue) and stars (yellow) are shown) and a virtual
slit for extracting a rotation curve. The slit width $d$ is
indicated by the red lines. A coordinate system as used in the
text (see Sect. 2.1) is shown.} \label{RCextraktor1}
\end{center}
\end{figure}

To follow the procedure applied in observations of distant, small,
and faint galaxies as closely as possible we define as a first
step a slit as shown in Fig. \ref{RCextraktor1}. Again, as in
paper I, we choose large relative slit widths as they are used in
observations of intermediate and high redshift galaxies (e.g.
B\"ohm et al. \cite{Boehm}; Nakamura et al. \cite{Nakamura}). As a
next step we project all stellar and gas particles of the
N-body/SPH simulations onto a cartesian, uniform grid, with a
spacing corresponding to typical spatial resolutions of the
observations of intermediate redshift galaxies, for example, in
the case of FORS1 the pixel scale of 0.2\arcsec, which corresponds
to $\sim 1.1$ kpc for a redshift of z=0.4. Subsequently we derive
a stellar blue band luminosity distribution $I(x,z)$ (the
coordinate system is chosen as indicated in Fig.
\ref{RCextraktor1}) by assuming a stellar mass-to-light ratio of
1.2. This is the mean obtained using different star formation
histories in the redshift range $z=0.5-1.4$ in Dickinson et al.
(2003). Then we calculate a luminosity-weighted velocity field
$V_{\rm{w}}$ as:
\begin{equation}
V_{\rm{w}}(x,z)=V(x,z) \times I(x,z),
\end{equation}
where $V(x,z)$ is the projected, 2D velocity field of the gas.
This weighting assures that brighter regions contribute more to
the velocity shift. To simulate seeing effects on our
velocity-field measurements, a convolution with a Gaussian point
spread function (PSF) was adopted on the luminosity-weighted
velocity field with an FWHM according to typical observational
conditions:
\begin{equation}
V_{\rm{w}}(x,z) \stackrel{PSF}{\longrightarrow} V_{\rm{wc}}(x,z).
\end{equation}
The FWHM of the seeing in observations is typically of the order
of 0.5"-1.2" (see e.g. J\"ager et al. \cite{Jaeger}). For our
simulations we adopt a mean value of 0.8" that corresponds, for
example, to 4.3 kpc for a source at redshift z=0.4.

\begin{figure}
\begin{center}
\includegraphics[width=\columnwidth]{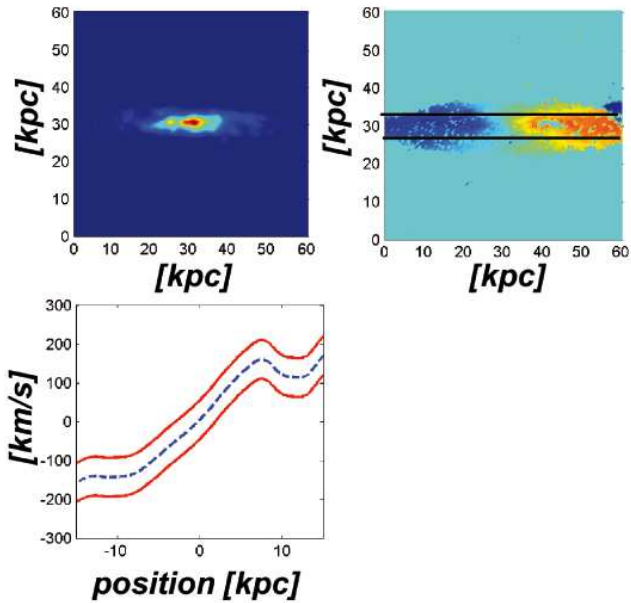}
\caption{Top left: luminosity field derived from the stellar-mass
distribution; top right: the unsmoothed and unweighted velocity
field of the gas. A virtual slit is indicated and a small part of
the gaseous disc of the second galaxy is visible on the right hand
side. Bottom left: the extracted RC for the modelled galaxy
(dashed line) with 1$\sigma$ errors (solid lines). For this image
series we use galaxy A in an early stage of an interaction
(unequal mass merger).} \label{AMBM50}
\end{center}
\end{figure}

Finally, we compute the mean velocity at a certain galactocentric
radius $r$, by averaging over all grid points within the slit
perpendicular to the spatial axis. Hence, the velocity extracted
for the RC at radius $r$ reads:
\begin{equation}
V_{\rm{RC}}(r) = \sum_{i=1}^N
\frac{V_{\rm{wc}}(x,z_i)}{I_{\rm{c}}(x,z_i)}.
\end{equation}
Here $x=r$ is the apparent galactocentric radius along the spatial
axis, and the sum runs over all grid points that lie within the
defined slit. For all RCs presented in this work, we place the
virtual slit along the major axis of the investigated galaxy. The
normalisation with the convolved luminosity distribution
$I_{\rm{c}}(x,z)$ is necessary to ensure that the rotational
velocity $V(r)$ of a constant velocity field would not be changed
by the weighting and the seeing. This approach allows a direct
comparison with 2D observations. In Fig. \ref{AMBM50} we show the
luminosity field derived from the stellar mass distribution, the
unsmoothed and unweighted velocity field of the gas, and finally
the extracted RC for one of our modelled galaxies in an early
stage of an interaction (unequal mass merger).

\begin{figure}
\begin{center}
\includegraphics[width=\columnwidth]{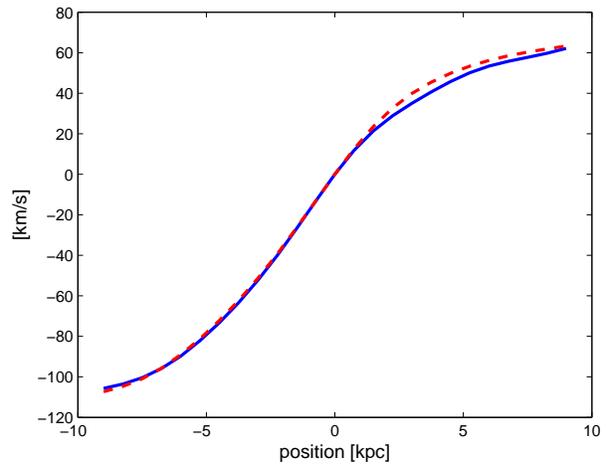}
\caption{Rotation curve extracted from one of our modelled systems
along the major axis using two different assumptions for the
mass-to-light ratio. The dashed line corresponds to a constant
mass-to-light ratio, while the solid line was extracted using a
linear decreasing M/L with a slope of 0.1 kpc$^{-1}$}
\label{ml_grad}
\end{center}
\end{figure}

We emphasise that the B-band mass-to-light ratio (M/L) is in
general not constant for typical spiral galaxies but a complex
function of radius (e.g. Verheijen et al. 2005). Therefore we
extracted rotation curves by assuming different radial gradients
in the mass-to-light ratio finding that the overall shape of the
RC, which we investigate in this paper, is not sensitive to the
specific choice of M/L. In Fig. \ref{ml_grad} we show RCs
extracted from one of our modelled systems using two different
assumptions for the mass-to-light ratio. In one case we used a
constant mass-to-light ratio while in the other case we extracted
the RC using a linearly decreasing M/L with a slope of 0.1
kpc$^{-1}$. Some quantitative differences are visible, but the
shape of the RC is not affected.

\subsection{The spatial alignment and impact parameters}
\begin{figure*}
\begin{center}
\includegraphics[width=\textwidth]{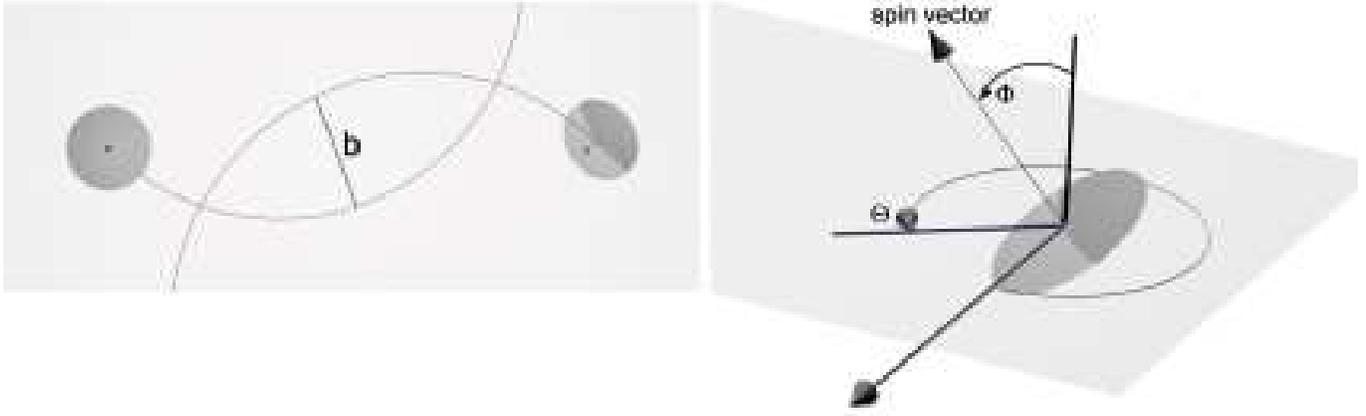}
\caption{Sketch to illustrate the parameters used to define the
interaction geometry. Figure from Kapferer et al. (2005). See
also Duc et al. (2000) for further descriptions.} \label{geom}
\end{center}
\end{figure*}

We follow the notation introduced by Duc et al. (2000) to describe
the interaction geometry. Parameter {\bf b} corresponds to the
minimum separation of the galaxies' trajectories, as if they were
point masses on Keplerian orbits (see Fig. \ref{geom}). The two
angles, $\Theta$ and $\Phi$, that define the spatial orientation
of the disc are described in Fig. \ref{geom}.

For the complete sample, we selected the alignments in such a way
as to cover as many geometries as possible, including minor and
major mergers and fly-bys (achieved by increasing the minimum
separation). The appearance of intrinsic RC distortions will of
course not only depend on the interaction geometry but also on the
angle under which the galaxy is observed (see Sect. 3.4). The
interaction geometries we use for this work are listed in Table
\ref{sims}, which represents a small sub-sample of the large
sample introduced by Kapferer et al. (2005).

\begin{table}
\caption[]{Interaction parameters for the simulations. C2: the two
interacting model galaxies; C7: minimum separation [kpc]; C8:
initial relative velocities of the two galaxies [km/s].}
\begin{center}
\begin{tabular}{c c c c c c c c}
\hline \hline Simulation & C2 & $\Phi_1$ & $\Theta_1$ & $\Phi_2$ &
$\Theta_2$ & C7 & C8 \cr \hline 1 & A-B & 0 & 0 & 0 & 0 & 5 & 250
\cr 2 & B-B & 0 & 0 & 0 & 0 & 5 & 120 \cr 3 & B-B & 0 & 0 & 0 & 0
& 50 & 120\cr 4 & B-B & 0 & 0 & 0 & 0 & 25 & 120\cr 5 & A-B & 0 &
0 & 180 & 0 & 5 & 250 \cr 6 & A-B & 0 & 0 & 0 & 0 & 25 & 250\cr 7
& A-B & 0 & 0 & 0 & 0 & 50 & 250\cr 8 & A-B & 0 & 0 & 90 & 0 & 5 &
250\cr 9 & B-B & 0 & 0 & 90 & 0 & 5 & 120\cr 10 & B-B & 0 & 0 &
180 & 0 & 5 & 120\cr \hline
\end{tabular}
\label{sims}
\end{center}
\end{table}

\section{Results}
In the following subsections we study the kinematic distortions
caused by unequal mass mergers, equal mass mergers, and fly-bys
and investigate the dependence of their appearance on the viewing
angle.

\subsection{Rotation curve distortions caused by unequal mass mergers}

\begin{figure}
\begin{center}
\includegraphics[width=\columnwidth]{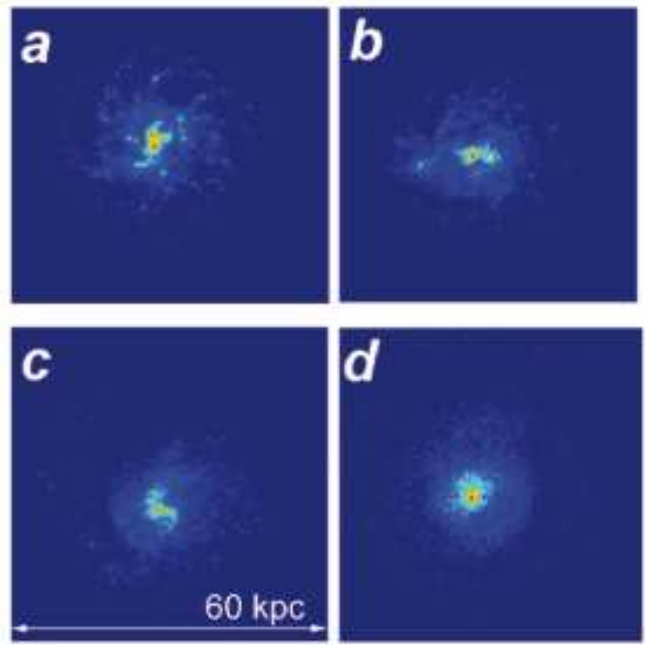}
\caption{Gaussian smoothed images of the luminosity field of the
galaxies, derived from the stellar mass distribution and with the
assumption of a constant mass-to-light ratio. The images (a) to
(d) correspond to the images in Fig. \ref{time_geo} but observed
under an inclination angle of 10$^\circ$. The sequence shows an
unequal mass merger, where image (a) represents 0.5 Gyr, (b) 0.6
Gyr, (c) 1 Gyr, and (d) 2 Gyr of evolution. More details about the
smoothing are listed in Sect. 2.1.} \label{rc_evo2d}
\end{center}
\end{figure}

To investigate the effects of unequal mass mergers on the RC of
the massive partner in the interaction, we use simulation 1 (see
Table \ref{sims}), which is a co-rotating coplanar collision of
model galaxy A with model galaxy B. Figure \ref{rc_evo2d} shows
the Gaussian smoothed images of the luminosity field of the
galaxies, derived from the stellar mass distribution and with the
assumption of a constant mass-to-light ratio (see Sect. 2.1). For
the spatial resolution of the images, we chose 0.1"/pixel (0.54
kpc for z=0.4), which corresponds to the optical resolution FWHM
of the ACS/WFC instrument on the Hubble Space Telescope (HST). It
is obvious from these images that minor mergers are not always
easy to detect in direct imaging, especially for intermediate and
high redshift galaxies. In the early stages of an interaction, it
is often possible to resolve the galaxy pairs, so the interaction
is clearly visible. Later on it is more difficult to distinguish
between an isolated galaxy and a late stage of interaction.

\begin{figure}
\begin{center}
\includegraphics[width=\columnwidth]{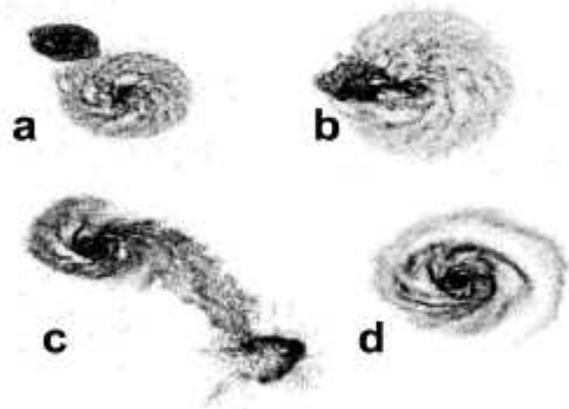}
\caption{Sequence of images corresponding to different time steps
in the merging process, with only the gas shown. The evolution of
the system is calculated over 5 Gyr. Here (a) represents 0.5 Gyr,
(b) 0.6 Gyr, (c) 1 Gyr, and (d) 2 Gyr of evolution.}
\label{time_geo}
\end{center}
\end{figure}

In Fig. \ref{time_geo} we show a sequence of images corresponding
to different time steps in the merging process, where only the gas
is shown. Note that the system is always viewed nearly face-on in
this figure, while the RC is extracted from the system observed
under an inclination angle of $i=80^{\circ}$ (edge-on is defined
by $i=90^{\circ}$). Chart (a) represents 0.5 Gyr, (b) 0.6 Gyr, (c)
1 Gyr, and (d) 2 Gyr of evolution. In chart (a) in Fig.
\ref{time_geo}, the galaxy is shown shortly ($\sim$ 100 Myr)
before the first encounter of the two gas discs, i.e. the moment,
in which the two gas discs overlap completely for the first time.
In the next chart (b) the two galaxies have their first encounter,
whereas charts (c) and (d) show the galaxies 0.4 Gyr and 1.4 Gyr
later, respectively. The massive partner (galaxy A) shows the
typical features of merging events such as tidal tails and
bridges. After 1.5 Gyr the perturbation in the gas distribution is
still visible in a pronounced spiral arm, as shown in Fig.
\ref{time_geo} (d) on the right hand side.

\begin{figure}
\begin{center}
\includegraphics[width=\columnwidth]{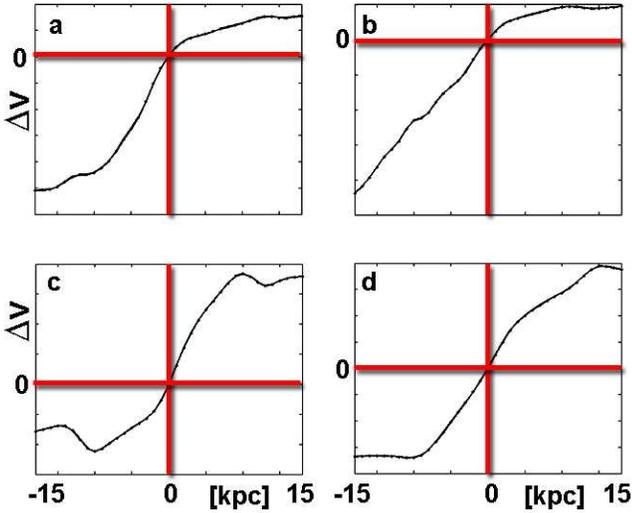}
\caption{Rotation curves for an unequal mass merger (galaxies A
and B) for different time steps, corresponding to the time series
shown in Fig. \ref{time_geo} but observed under an inclination of
i=80$^\circ$. The red lines always indicate the centre of galaxy A
to highlight asymmetries in the RCs.} \label{rc_evo}
\end{center}
\end{figure}

In Fig. \ref{rc_evo} the corresponding RCs are shown. The RCs were
extracted as described in Sect. 2.1 for model galaxy A. Before the
first encounter, the RC has an undisturbed shape. As galaxy B
comes closer, galaxy A feels an acceleration towards the common
centre of mass. This acceleration affects the velocity field in
such a way that relative velocities with respect to the centre of
galaxy A are enhanced towards the approaching galaxy B and
weakened on the opposite side. In other words, the kinematic
centre of the galaxy is shifted with respect to the geometrical
centre of the gaseous disc. The RC of the merging-phase (b) shows
strong signatures of the first encounter in its asymmetric shape.
The incorporated galaxy B with its velocity component along the
line of sight causes the acceleration on the left hand side of the
RC. The right hand side of the RC is not affected yet. This part
of the RC is not distorted and does resemble the 'classical'
rotation curve shape rising in the inner part and turning over to
a flat regime. The third RC corresponds to a later evolutionary
step (400 Myr after RC (a)), where the less massive galaxy has
passed galaxy A and left behind a disturbed velocity field.
Although the RC has become almost symmetric again, it shows signs
of interaction, especially the pronounced bumps in the outer parts
($\pm\sim$ 10 kpc). The last RC (d) shows the system 1.4 Gyr after
the first encounter; note that galaxy B is no longer visible in
the direct vicinity. Indeed the system has almost relaxed and
nearly no signs of the interaction are left.

In order to quantify the degree of asymmetry of the RC shape, we
follow an approach used in Dale et al. (\cite{Dale}). We measure
the area between the kinematically folded approaching and receding
halves and divide it by the average area under the RC:

\begin{eqnarray}\label{assymeq}
\textrm{Asymmetry}&=&\sum
\frac{||V(r)|-|V(-r)||}{\sqrt{\sigma^2(r) +\sigma^2(-r)}}
\nonumber\\ &\times& \left[\frac{1}{2} \sum
\frac{|V(r)|+|V(-r)|}{\sqrt{\sigma^2(r)+\sigma^2(-r)}}\right]^{-1}.
\end{eqnarray}

Here $\sigma(r)$ is the uncertainty of the rotational velocity at
position $r$. We adopted a value of 20 km/s, which is typically
the minimum uncertainty in observations of distant galaxies due to
the limited spectral resolution. In Table \ref{assymtable 1} we
list the asymmetry measure for different instances under different
viewing angles for the unequal mass merger of simulation 1. The
asymmetry of the RC increases with time as the two galaxies
approach each other. After 0.5 Gyr a significant asymmetry is
present along all lines-of-sight. The asymmetry reaches a maximum
after 0.7 Gyr of evolution and subsequently falls to the value it
had before the first encounter. The small drop in the asymmetry
measure after 0.9 Gyr corresponds to the time between the first
and second encounter. Note that the asymmetry strongly depends on
the viewing angle at a certain instant. We will discuss this issue
in Sect. \ref{view}. The small companion galaxy B, on the
contrary, already loses most of its gas during the first pass
through galaxy A and finally merges completely with the larger
galaxy; hence, the distortions in the velocity field are so severe
that we cannot make a more detailed investigation.

\begin{table}
\caption[]{Evolution of the asymmetry parameter$^1$ for unequal
mass mergers (simulations 1, 5, and 8) under different
lines-of-sight (angle $\Phi$ is measured with respect to an
arbitrarily chosen line-of-sight).}\label{assymtable 1}
\begin{center}
\begin{tabular}{l || c c c c c c}
Sim. 1: \cr \hline time & 100 & 500 & 700 & 900 & 1100 & 2000 \cr
$\Phi$ & [Myr] & [Myr] & [Myr] & [Myr] & [Myr] & [Myr]
\cr\hline\hline 0$^\circ$ & 9\% & 54\% & 103\% & 20\% & 100\% &
12\%\cr 45$^\circ$ & 10\% & 44\% & 120\% & 58\% & 103\% & 17\%\cr
90$^\circ$ & 7\% & 21\% & 112\% & 67\% & 89\% & 8\%\cr 135$^\circ$
& 8\% & 52\% & 45\% & 46\% & 51\% & 7\%\cr\hline
Sim. 5: \cr \hline time & 100 & 500 & 700 & 900 & 1100 & 2000 \cr
$\Phi$ & [Myr] & [Myr] & [Myr] & [Myr] & [Myr] & [Myr]
\cr\hline\hline 0$^\circ$ & 10\% & 50\% & 48\% & 20\% & 45\% &
6\%\cr 45$^\circ$ & 10\% & 11\% & 51\% & 13\% & 12\% & 18\%\cr
90$^\circ$ & 12\% & 70\% & 38\% & 15\% & 33\% & 12\%\cr 135$^\circ$
& 9\% & 61\% & 12\% & 9\% & 31\% & 17\%\cr\hline
Sim. 8: \cr \hline time & 100 & 500 & 700 & 900 & 1100 & 2000 \cr
$\Phi$ & [Myr] & [Myr] & [Myr] & [Myr] & [Myr] & [Myr]
\cr\hline\hline 0$^\circ$ & 11\% & 21\% & 122\% & 44\% & 19\% &
13\%\cr 45$^\circ$ & 9\% & 93\% & 83\% & 38\% & 26\% & 17\%\cr
90$^\circ$ & 13\% & 74\% & 40\% & 13\% & 28\% & 20\%\cr 135$^\circ$
& 10\% & 110\% & 38\% & 10\% & 37\% & 7\%\cr\hline
\end{tabular}
\end{center}
$^{1}$....asymmetry measure as defined by Dale et al. 2001, see
Eq. \ref{assymeq}
\end{table}

From both theoretical and observational studies, it is expected
that kinematic disturbances fade within a few rotation cycles
($\le$1 Gyr) and therefore reflect the recent interaction history
(e.g. Rubin et al. \cite{Rubin}; Dale et al. 2001). Our results
from this and the subsequent sections confirm this statement
numerically within the investigated interaction scenarios. The
damping of the disturbances is listed quantitatively in Table
\ref{assymtable 1}. 1.4 Gyr after the first encounter, i.e. after
2 Gyr of evolution, the asymmetry measure reaches a value close to
the one after 0.1 Gyr of evolution, i.e. before the first
encounter.

\begin{figure}
\begin{center}
\includegraphics[width=\columnwidth]{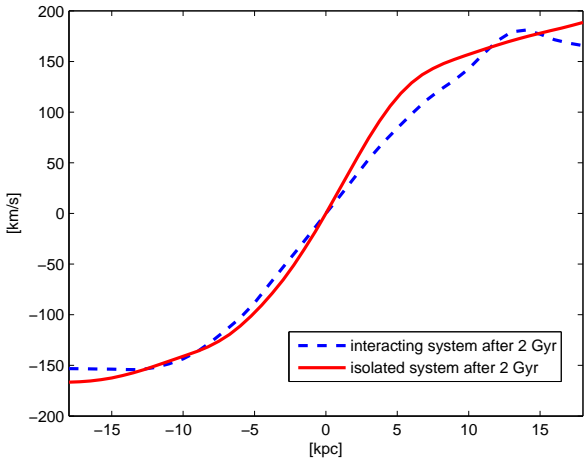}
\caption{Rotation curve of model galaxy A (blue, dashed line) 1.4
Gyr after the first encounter with model galaxy B, i.e. after 2
Gyr of evolution, as compared to the rotation curve of the
isolated model galaxy A (red, solid line) after 2 Gyr of isolated
evolution.} \label{vorher_nachher}
\end{center}
\end{figure}

In Fig. \ref{vorher_nachher} the RC for model galaxy A 1.4 Gyr
after the first encounter, i.e. after 2 Gyr of evolution is
compared to the RC of the isolated model galaxy A after 2 Gyr of
isolated evolution. The shape of the RC for model galaxy A
resembles $\sim$1 Gyr after the first encounter the one of an
isolated system, so the properties derived from it would not
differ significantly within given errors. The only significant
difference is the slope of the inner part of the RC. For
intermediate and high-redshift galaxies, small bumps as in the
right section of the red line would not be observable.

We study effects caused by collisions of \textit{counter-rotating}
discs using simulation 5 (see Table \ref{sims}). The shape of the
RCs do differ from those of the co-rotating case, as different
velocity fields are superposed. However, there are also velocity
components due to the movement of galaxy B on its orbit that is
approaching galaxy A. These are the same for both configurations
and are dominating for the unequal mass merger, which we
investigate here. Connected to the relative trajectories of the
galaxies is also a transfer of angular momentum, leading to
increasing rotational velocities (Heyl et al. 1996); therefore,
the shape of the RC of galaxy A during the interaction shows only
a small dependence on the rotational direction of galaxy B.

\subsection{Rotation curve distortions caused by equal mass
mergers}\label{sec_equal}

\begin{figure}
\begin{center}
\includegraphics[width=\columnwidth]{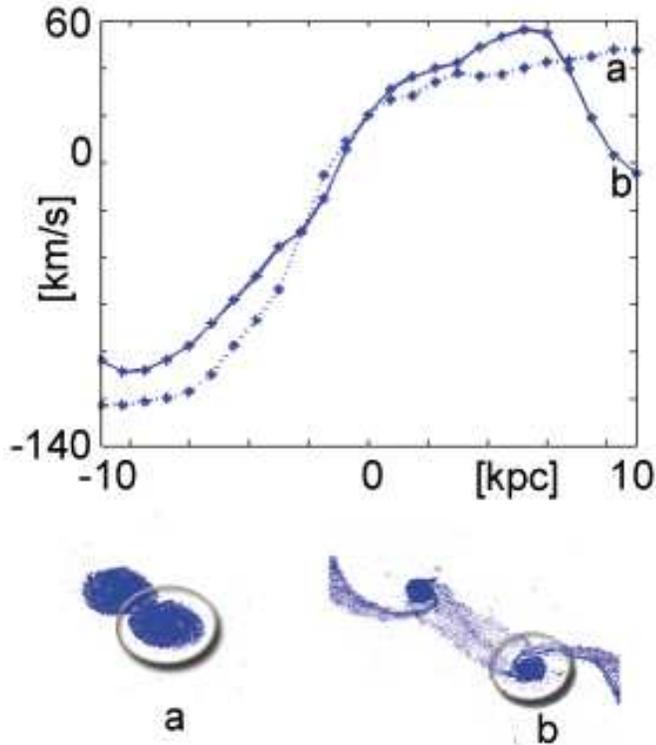}
\caption{Rotation curves for an equal mass merger for different
time steps corresponding to the interaction geometry illustrated
in the lower panel. The galaxy from which the RC was extracted is
highlighted. Here panel (a) shows the system after 1.1 Gyr and
panel (b) after 2 Gyr of evolution.} \label{equal1}
\end{center}
\end{figure}

In a next step we investigate the effects of equal mass mergers on
the RCs of the interacting system. For this purpose we use
simulation 2 (see Table \ref{sims}), which is a co-rotating
coplanar collision of two model galaxies B. Equal mass mergers
massively disturb the velocity fields of the involved galaxies.
Typically, the gaseous disc is almost completely destroyed in the
merging process, and the orbits of the stars are randomized,
leading to the formation of an elliptical galaxy. This result has
been shown in many previous studies (e.g. Barnes \& Hernquist
1991; Cretton et al. 2001; Naab \& Burkert 2003; Bournaud et al.
2005). We also find this formation of an elliptical galaxy in our
simulations; however, as this issue was investigated
comprehensively by other authors, we do not discuss this in more
detail in the present work. Instead, we chose two snapshots for
which the gaseous disc is still observable and extracted the
rotation curves for one of the interacting galaxies. The result is
shown in Fig. \ref{equal1} together with the corresponding
interaction geometry. Note that in this figure the geometry is
again illustrated as observed from nearly face-on, while the RC is
extracted from the system observed under an inclination angle of
$i=80^{\circ}$, as in the previous section. Panel (a) shows the
system after 1.1 Gyr and panel (b) after 2 Gyr of evolution. The
first encounter of the two gas discs happens after 1.2 Gyr of
evolution, i.e. 0.1 Gyr after the snapshot shown in panel (a).

\begin{figure}
\begin{center}
\includegraphics[width=\columnwidth]{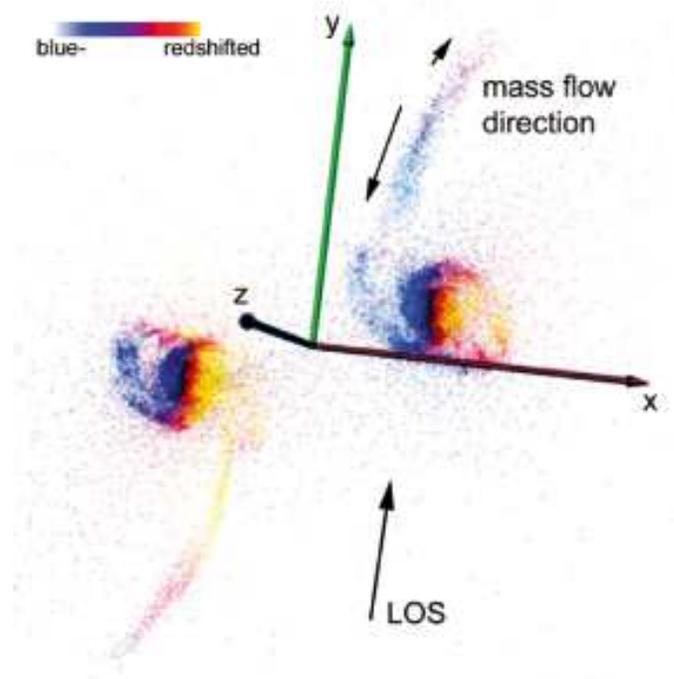}
\caption{Sketch to illustrate the decrease in the rotational
velocities in the outer parts of the RCs that can be caused by gas
from the tidal tail, which has small radial velocity components at
this projected position and is covered by the slit.} \label{jucy}
\end{center}
\end{figure}

Before the first encounter, the RC of the investigated galaxy is
only slightly disturbed but shows an asymmetric shape. As in the
previous section, the kinematic centre of the galaxy is shifted
with respect to the geometrical centre, i.e. the maximum of the
surface mass density of the gaseous disc, due to the influence of
the approaching galaxy. As the two discs pass through each other,
tidal tails and the bridges are formed. These features can be seen
in panel (b), where the distribution of the gas is shown. An RC
extracted from the highlighted galaxy does not show a symmetric
shape either. Again the merging process introduces a strong
acceleration on the right hand side in the galaxy, leading to the
formation of a tidal tail. Therefore, the relative velocities on
the right hand side decrease with respect to the centre, while the
left hand side shows higher relative velocities, leading to the
asymmetric shape. The decrease in the rotational velocities in the
outermost parts on the right hand side of the RC emerges as gas
from the tidal tail is covered by the slit. The gas and the stars
of the tail falling back onto the rotating disc have a small
radial velocity component at this projected position, as
illustrated in Fig. \ref{jucy}. This possible change in the sign
of the velocity gradients in the presence of tidal tails has
already been found by Bournaud et al. (2004), who also points out
the strong sensitivity of this result on the viewing angle. (We
investigate this dependence on the viewing angle in Sect.
\ref{view}.) A significant decrease in the outer parts of the RC
(see Fig. \ref{equal1}) could indicate the presence of a massive
tidal tail as consequence of a recent major merger event. However,
this feature is not unique for tidal tails. As we will show in
Sect. \ref{view}, the viewing angle of the observation also plays
an important role (see also Bournaud et al. 2004).

\subsection{Rotation curve distortions caused by fly-bys}

\begin{figure}
\begin{center}
\includegraphics[width=\columnwidth]{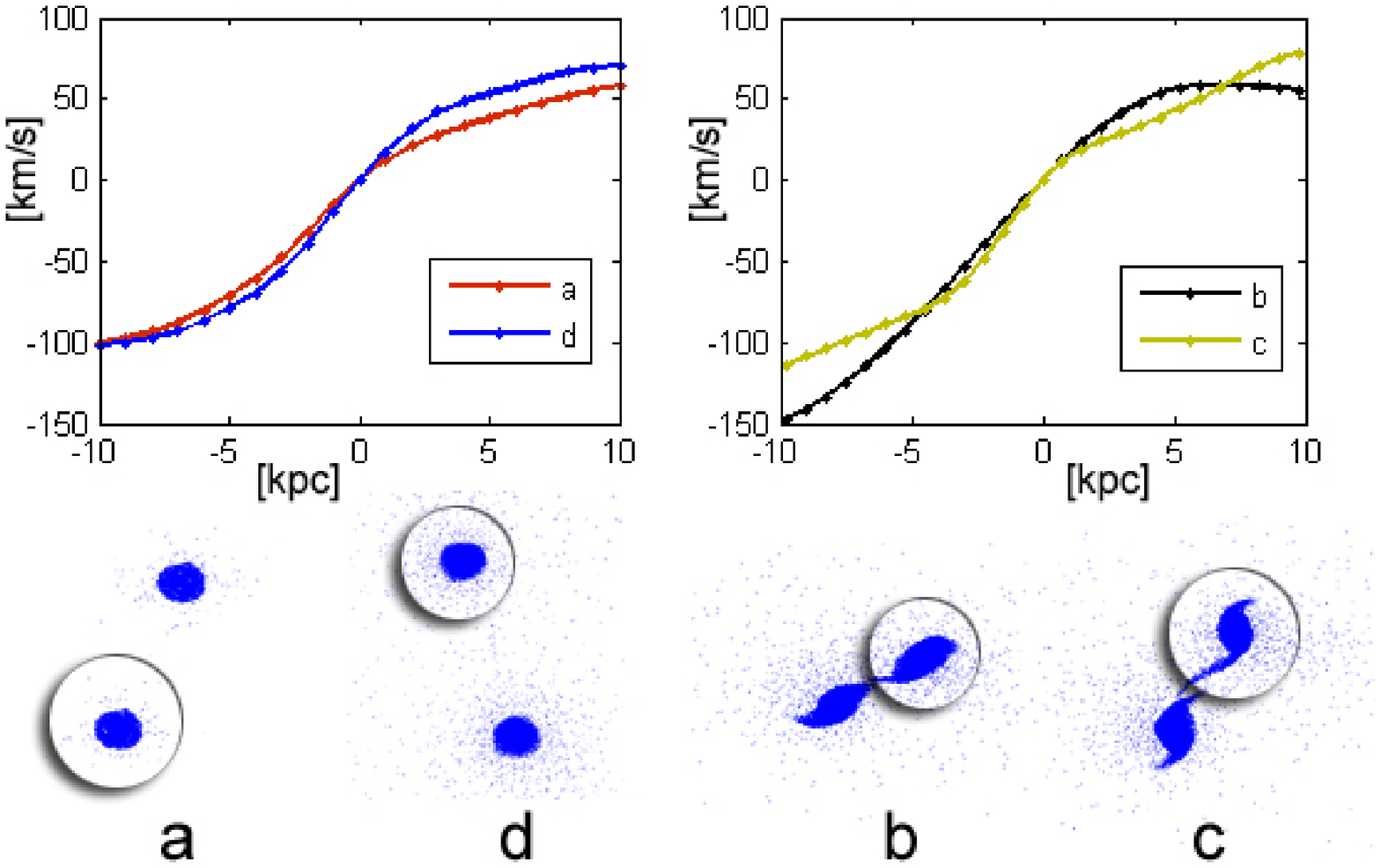}
\caption{Rotation curves for different snapshots of a fly-by of
two model galaxies B. In the lower panel the alignment of the
system and in the upper panel the corresponding rotation curves of
the highlighted galaxy are shown. Here (a) represents 0.5 Gyr, (b)
1.7 Gyr, (c) 2 Gyr, and (d) 3 Gyr of evolution.} \label{fly_by}
\end{center}
\end{figure}

In order to estimate the influence of fly-bys on the shape of RCs,
we use simulation 3 (see Table \ref{sims}), which has a minimum
separation of 50 kpc (at the first encounter). In Fig.
\ref{fly_by} the RC of the highlighted galaxy for different
snapshots is presented in the upper panel. In the lower panel the
geometry corresponding to the different time steps of the
interaction is shown. Panel (a) represents 0.5 Gyr, (b) 1.7 Gyr,
(c) 2 Gyr, and (d) 3 Gyr of evolution. The change in the velocity
field of the gaseous disc on short timescales is apparent in the
right panel. The geometry (b) represents the nearest encounter,
which is visible by a highly asymmetric shape. On the side
pointing towards the companion, the RC does not flatten, due to
the additional acceleration. On the opposite side, this feature is
not present, instead the RC flattens. After 300 Myr (c), the
velocity field does not show this asymmetry anymore, although the
overall shape does not resemble an undisturbed RC. In the left
panel we show the RC 0.5 Gyr after the beginning of the simulation
(a) and after 2.5 Gyr of evolution (d). Note that the galaxies are
spatially well-separated in both cases and that they would be
therefore considered as isolated systems especially at high
redshifts. The fly-by does not lead to persistent changes in the
velocity field, and consequently the RCs do not show significant
differences before and after the fly-by. The closest approach and
the formation of observable distortions happens after $\sim$ 2 Gyr
of evolution (panels (b) and (c)). After 3 Gyr of evolution, i.e.
1 Gyr after the appearance of the first distortions, the RC no
longer shows signs of interaction. Thus, the velocity field again
settles in this case to a relatively undisturbed equilibrium state
after $\sim$1 Gyr.

Again we quantify the degree of asymmetry using Eq. \ref{assymeq}
and present the results in Table \ref{assymtable 2}. We find an
increasing value for the asymmetry with time as the two galaxies
approach each other. After $\sim$ 2 Gyr of evolution, the two
galaxies come closest to each other. One Gyr later, the RC of the
investigated galaxies shows a relatively undisturbed form again
under most viewing angles, which again quantifies the settling
time of $\sim$ 1 Gyr. In this interaction scenario we also find
the strong dependence on the viewing angle in an interacting
system.

Decreasing the minimum separation to 25 kpc as in simulation 4
(see Table \ref{sims}) leads to slightly stronger distortions
during the closest approach of the two galaxies, both in the mass
distribution and in the kinematics. Nevertheless, the system
settles back to an equilibrium state on approximately the same
timescale as in the more distant fly-by discussed above ($\sim$ 1
Gyr). A fly-by of a less massive galaxy leads to lower asymmetries
in the rotation curve (see simulation 7 in Table \ref{assymtable
2}).

\begin{table*}
\caption[]{Evolution of the asymmetry parameter$^1$ for fly-bys
(simulations 3, 4, and 7) under different lines-of-sight (angle
$\Phi$ is measured with respect to an arbitrarily chosen
line-of-sight).}\label{assymtable 2}
\begin{center}
\begin{tabular}{l || c c c c c c c c}
Sim. 3: \cr \hline time & 700 & 900 & 1100 & 1300 & 1500 & 1700 &
2000 & 3000 \cr $\Phi$ & [Myr] & [Myr] & [Myr] & [Myr] & [Myr] &
[Myr] & [Myr] & [Myr] \cr\hline\hline 0$^\circ$ & 12\% & 25\% &
45\% & 49\% & 89\% & 99\% & 44\% & 14\% \cr 45$^\circ$ & 4\% &
19\% & 35\% & 59\% & 35\% & 30\% & 19\% & 9\%\cr 90$^\circ$ & 15\%
& 21\% & 15\% & 62\% & 16\% & 16\% & 17\% & 4\%\cr 135$^\circ$ &
20\% & 18\% & 21\% & 19\% & 97\% & 2\% & 4\% & 23\% \cr\hline
Sim. 4: \cr \hline time & 700 & 900 & 1100 & 1300 & 1500 & 1700 &
2000 & 3000 \cr $\Phi$ & [Myr] & [Myr] & [Myr] & [Myr] & [Myr] &
[Myr] & [Myr] & [Myr] \cr\hline\hline 0$^\circ$ & 16\% & 17\% &
17\% & 16\% & 37\% & 20\% & 15\% & 8\% \cr 45$^\circ$ & 25\% &
16\% & 31\% & 19\% & 16\% & 53\% & 16\% & 5\%\cr 90$^\circ$ & 26\%
& 17\% & 48\% & 22\% & 40\% & 20\% & 21\% & 10\%\cr 135$^\circ$ &
30\% & 10\% & 21\% & 24\% & 36\% & 90\% & 16\% & 20\% \cr\hline
Sim. 7: \cr \hline time & 700 & 900 & 1100 & 1300 & 1500 & 1700 &
2000 & 3000 \cr $\Phi$ & [Myr] & [Myr] & [Myr] & [Myr] & [Myr] &
[Myr] & [Myr] & [Myr] \cr\hline\hline 0$^\circ$ & 8\% & 15\% &
17\% & 6\% & 9\% & 12\% & 13\% & 5\% \cr 45$^\circ$ & 20\% &
8\% & 14\% & 13\% & 8\% & 15\% & 9\% & 10\%\cr 90$^\circ$ & 21\%
& 13\% & 16\% & 20\% & 13\% & 14\% & 7\% & 12\%\cr 135$^\circ$ &
17\% & 14\% & 12\% & 21\% & 15\% & 15\% & 6\% & 10\% \cr\hline
\end{tabular}
\end{center}
\begin{center}
$^{1}$....asymmetry measure as defined by Dale et al. 2001, see
Eq. \ref{assymeq}
\end{center}
\end{table*}

\subsection{Dependence of the results on the viewing
angle}\label{view}

\begin{figure}
\begin{center}
\includegraphics[width=\columnwidth]{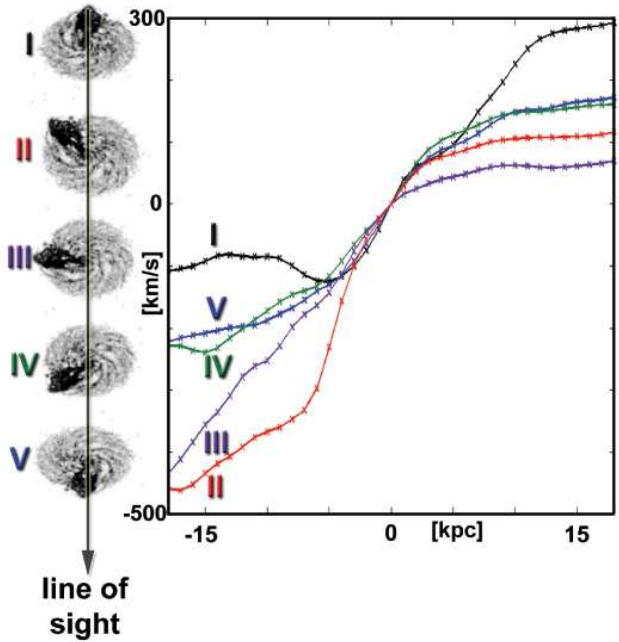}
\caption{Rotation curves for the first encounter of model galaxy A
and B seen nearly edge on (i=80$^\circ$) from different viewing
angles. The geometry is illustrated by the image series on the
left hand side. The snapshot corresponds to panel (b) in Fig.
\ref{time_geo} and therefore shows the system after 0.6 Gyr of
evolution.} \label{dreh_final}
\end{center}
\end{figure}

In an interacting system, two different velocity fields overlap.
Therefore, the shape of the RC, which is a 1D representation of
the full 3D velocity field, will significantly depend on the
viewing angle. This fact is shown in Fig. \ref{dreh_final} for the
RCs extracted from the first encounter of model galaxies A and B
seen nearly edge on from different viewing angles. The geometry is
illustrated by the image series on the left hand side of the
figure. Again this image series shows the system face-on, while
the RCs are extracted from the galaxy observed under an
inclination angle of $i=80^{\circ}$. We varied the azimuthal angle
in steps of 45$^{\circ}$. The differences in the RCs are striking
and are mainly caused by the velocity components of galaxy B on
its trajectory and its projections.

\begin{figure}
\begin{center}
\includegraphics[width=7cm]{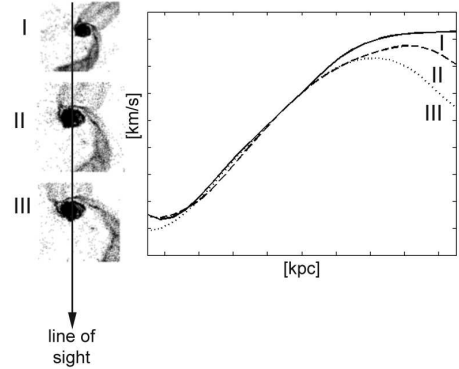}
\caption{Rotation curves for model galaxy A after an equal mass
merger from different viewing angles. The geometry is illustrated
by the image series on the left hand side. The snapshot
corresponds to panel (b) in Fig. \ref{equal1} and therefore shows
the system after 2 Gyr of evolution. The change in the sign of the
velocity gradient does not appear for all viewing angles.}
\label{tidaltails}
\end{center}
\end{figure}

As we have shown in Sect. \ref{sec_equal}, the presence of a tidal
tail can lead to a change in the sign of the velocity gradient,
which was also described by Bournaud et al. (\cite{Bournaud}).
However, this feature depends on the viewing angle, as in
long-slit spectra (and also in position-velocity diagrams obtained
from Fabry-Perot interferometry) we always look at only one (the
line-of-sight) component of a 3D vector field. If this component
is smaller at a certain position, because, for example, the gas
there has been decoupled from the general rotation of the disc by
tidal interaction, a decrease in the RC can be found. In Fig.
\ref{tidaltails} we show the interacting system with pronounced
tidal tails from different lines-of-sight. Note that even a flat
RC is possible in some configurations, although we found a change
in the sign of the velocity gradient for most lines-of-sight. We
also note that a prominent turn over in the RC is present in the
third instance shown in Fig. \ref{tidaltails}. A part of the tidal
tail is aligned with the line-of-sight. In their simulations for
such configurations Bournaud et al. (\cite{Bournaud}) always found
a change in the sign of the velocity gradient in their
position-velocity diagrams, consistent with our results shown in
Fig. \ref{tidaltails}. In addition to the line-of-sight, the time
evolution of the tidal tail also plays an important role in the
question of whether the turn over in the RC can be observed.
Hibbard \& Mihos (\cite{Hibbard}) showed that the kinematics of
the tidal material depends on the radius, as well as on the time
evolution. A decreasing outer part in the RC at a certain instant
and under a certain line-of-sight can be interpreted, as pointed
out in Fig. \ref{jucy}. However, this feature is not unique, as we
mentioned before and a tidal tail is just one possible
explanation.

As there is an infinite number of viewing angles to an infinite
number of different interaction geometries, it is impossible to
relate a specific observed RC distortion unambiguously to a
certain interaction scenario without further information e.g. from
high-resolution direct imaging. The specific shape of the RC will
always depend on both the geometry of the interaction and the
viewing angle. The situation is further complicated, as the
dynamical evolution of the system is highly non-linear. However,
these interaction-induced distortions have in common that the RCs
become asymmetric, show significantly rising or falling profiles
on the side pointing towards the other galaxy, and show pronounced
bumps depending on the viewing angle (cf. Figs. \ref{rc_evo} and
\ref{dreh_final}). We found no observational biases for isolated
galaxies that could produce such pronounced features (see paper I
for details). Most observational effects change the RC of a galaxy
in a symmetric way (e.g. relative large slit widths, inclination).
Spiral arms typically cause fluctuations of a few tens of km/s
that are superposed on the smooth rotation curve of the galaxy
(see e.g. Sofue and Rubin, 2001 and paper I). Thus spiral arms can
only account for smaller distortions but not for high asymmetries,
as present in the present paper. Of course it is important to
stress that galaxies in the universe are never isolated. The
dynamical history of a galaxy will always influence the kinematics
of the different components of galaxies. Not all asymmetries are
caused by clearly observable mergers/interactions (Emergreen 2005
and references therein), although distinct asymmetries are often
observed in interacting systems. Very recently disturbed isolated
galaxies were investigated to search for indicators of
interactions with dark galaxies, i.e. DM halos without any
baryonic matter (Karachentsev et al. 2006).

\section{Implications of the results for the observational properties of cluster galaxies}
If the results, which were found in the previous sections for a
restricted sample of interaction geometries, hold more generally,
then they have an important implication for observational
properties of cluster galaxy populations. First of all, stable
regular discs show the classical RC, rising in the inner part and
turning over to a flat regime. Interactions as mergers or fly-bys
introduce instabilities to the gas, which lead to an enhanced star
formation rate (e.g. Kapferer et al., 2005), provided the galaxy
has not lost large fractions of its gas e.g. by ram-pressure
stripping. As shown in the previous sections, these interactions
also cause kinematic distortions. After the interaction, both the
star formation and the kinematic distortions again settle back
into a stable equilibrium state. Our results therefore suggest
that enhanced star formation in apparently isolated galaxies can
be triggered by non-equal mass mergers. In direct imaging,
especially for high-redshift galaxies, those interactions are
hardly visible. Additional investigation of the velocity field
would help to reveal the dynamical state of the system. This
implies that distortions in RCs can give evidence of the star
formation of a galaxy and vice versa. As consequences of
interactions being even longer observable in the velocity field
than in direct imaging, these investigations can help to extract
clues for the star formation history of a galaxy.

\begin{figure*}
\begin{center}
\includegraphics[width=\textwidth]{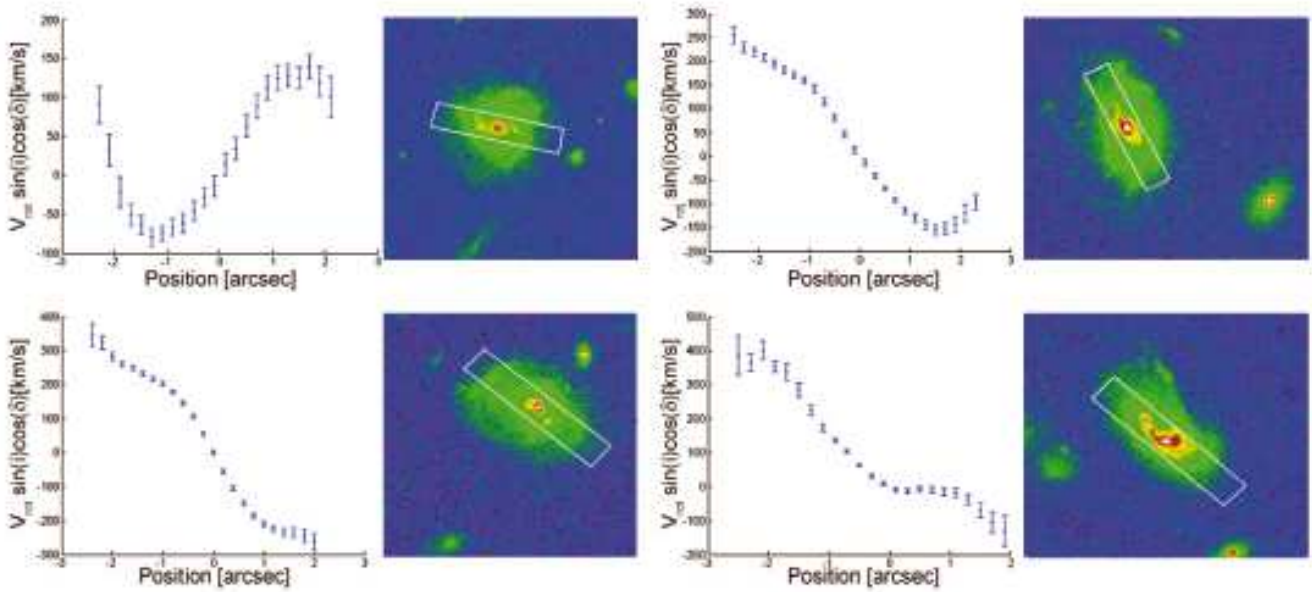}
\caption{ACS images taken in the F814 filter and rotation curves
from the VLT/FORS spectroscopy of four galaxies in the distant
cluster MS 0451.5-0305 at $z$=0.53. Tentatively, we find evidence
for, e.g., tidal streams (upper panel left, cf. Fig.8) or a recent
close fly-by/merger (lower panel right, cf. Fig. 10). Note that
the rotation curves give the observed values of the rotational
velocity V$_{\rm rot}$; i.e. they are not corrected for
inclination or offset angles ($\delta$) between the slit
directions and the galaxies' major axes. The indicated rectangles
correspond to the extensions of the derived RCs.} \label{obs}
\end{center}
\end{figure*}

We now discuss briefly to what extent observations of distant
cluster galaxies can be interpreted by simulations as presented
here. To that purpose, we show thumbnails and rotation curves of
four galaxies in Fig. \ref{obs} as examples. They are members of
the cluster MS\,0451.5--0305 at a mean redshift of $z = 0.53$,
which was recently imaged by Ellis et al. using the ACS camera
onboard HST in the F814W filter. The VLT/FORS spectra were
obtained as part of a larger program to investigate galaxy
transformations as described by Ziegler et al. (\cite{Ziegler})
(see also J\"ager et al., \cite{Jaeger}). Tilted MXU slitlets were
placed along the photometric major axis as determined on
\textit{groundbased} images (VLT/FORS) since no full coverage of
the cluster field by HST was available yet. The slitlets with
0.9\arcsec\ width are actually much longer than indicated on the
thumbnails. The position-velocity diagrams (still unpublished)
present the measured shifts of emission lines using the method
introduced by B\"ohm et al. (\cite{Boehm}) and J\"ager et al.
(\cite{Jaeger}) so that the displayed velocities are not corrected
for the galaxy inclination and the effects caused by the
observations. Note, however, that our V$_{\rm max}$ measurements
fully accounted for geometrical (e.g. inclination) and
observational effects (like seeing).

Two of the galaxies have rather regular structures and rotation
curves, while the other two spirals are clearly distorted. They
exhibit features that are very similar to what is seen in our
simulations.

\section{Summary and conclusions}
In this work we have investigated the rotation curves of
interacting spiral galaxies using N-body/SPH simulations with
included prescriptions for cooling, star formation, and stellar
feedback. We focussed on the questions of which distortions of the
RCs can arise from mergers (in unequal mass mergers the mass ratio
of our galaxies is 8:1) and fly-bys (25 - 50 kpc), how long these
disturbances survive, and what is their dependence on the viewing
angles. To summarise and conclude:

\bit \item Galaxy-galaxy mergers significantly disturb the
velocity fields and hence the RCs of the interacting galaxies.
These distortions commonly introduce asymmetries in the RCs with
significantly rising or falling profiles on the side pointing
towards the companion galaxy and pronounced bumps. These
distortions are also clearly observable for distant galaxies.

\item The specific shape of the RC depends on the mass ratio of
the interacting galaxies, the geometry of the interaction, and the
viewing angle. Therefore, without any additional information on
the system, it is impossible to relate a specific, observed RC
distortion unambiguously to a certain interaction scenario.
High-resolution direct imaging and 2D velocity fields allow
further insight into the dynamical state of the system.

\item We used a quantitative measure for the RC asymmetry and find
for given interaction scenarios that the disturbed velocity field
settles relatively fast ($\sim$1 Gyr) to a stable undistorted
equilibrium state after an unequal mass merger or a fly-by. This
numerically confirms results from both previous theoretical and
observational studies. Therefore, the shapes of the RCs before and
after such an interaction are very similar, especially as there
are no persistent distortions visible in the RCs. Investigations
of the number of interactions as a function of redshift including
velocity fields could be strongly biased by this effect.

\item Tidal tails, commonly formed by equal mass mergers, can be
visible in the rotation curves as a decrease in the rotational
velocities, provided the gas with a small radial velocity
component lies within the slit. However, we showed that this
change in the sign of the velocity gradient is not unique, and a
tidal tail is just one of the possible explanations. Additionally,
the viewing angle on the system plays an important role in the
question of whether the turn over in the RC can be observed.

\item Two effects contribute to the distortions in the velocity
field of an interacting system, starting with the superposition of
the two individual velocity fields that are modified by the
disturbed gravitational potential. Also, the velocity components
from the trajectories of the galaxies add to the resulting
velocity field, and the relative orbits cause an angular momentum
transfer (see also Heyl et al. 1996).

\eit

\section*{Acknowledgements}

The authors would like to thank Volker Springel for providing them
with GADGET2 and his initial-conditions generator. We are grateful
to the anonymous referee for criticism that helped to improve the
paper. The authors acknowledge the Austrian Science Foundation
(FWF) through grant number P15868, the UniInfrastrukturprogramm
2004 des bm:bwk Forschungsprojekt Konsortium Hochleistungsrechnen,
the German Science Foundation (DFG) through Grant number Zi
663/6-1, and the Volkswagen Foundation (I/76 520). In addition,
the authors acknowledge the Deutsches Zentrum f\"ur Luft und
Raumfahrt through grant 50 OR 0301, the ESO Mobilit\"atsstipendien
des bm:bwk (Austria), and the Tiroler Wissenschaftsfonds
(Gef\"ordert aus Mitteln des vom Land Tirol eingerichteten
Wissenschaftsfonds).


\begin{thebibliography}{} 

\bibitem[2005]{Bamford} Bamford, S.~P.,
Milvang-Jensen, B., Arag{\'o}n-Salamanca, A., \& Simard, L.\ 2005,
\mnras, 361, 109

\bibitem[1991]{Barnes} Barnes, J.~E., \&
Hernquist, L.~E. 1991, \apjl, 370, L65

\bibitem[1999]{Barton} Barton, E.~J., Bromley,
B.~C., \& Geller, M.~J., 1999, \apjl, 511, L25

\bibitem[2004]{Boehm}
B{\" o}hm, A., et al.\ 2004, A\&A, 420, 97

\bibitem[2004]{Bournaud} Bournaud, F., Duc,
P.-A., Amram, P., Combes, F., \& Gach, J.-L. 2004, \aap, 425, 813

\bibitem[2005]{Bournaud2} Bournaud, F., Jog,
C.~J., \& Combes, F.\ 2005, \aap, 437, 69

\bibitem[2006]{Chemin}
Chemin, L., et al.\ 2006, \mnras, 366, 812

\bibitem[1999]{Conselice} Conselice,
C.~J., \& Gallagher, J.~S. 1999, \aj, 117, 75

\bibitem[2001]{Cretton} Cretton, N., Naab, T.,
Rix, H.-W., \& Burkert, A. 2001, \apj, 554, 291

\bibitem[2001]{Dale} Dale, D.~A., Giovanelli,
R., Haynes, M.~P., Hardy, E., \& Campusano, L.~E. 2001, \aj, 121,
1886

\bibitem[2003]{Dickinson}
Dickinson, M., Papovich, C., Ferguson, H.~C., \& Budav{\' a}ri, T.
2003, ApJ, 587, 25

\bibitem[1994]{Dressler} Dressler, A., Oemler,
A.~J., Butcher, H.~R., \& Gunn, J.~E. 1994, \apj, 430, 107

\bibitem[2000]{Duc}
Duc, P.-A., Brinks, E., Springel, V., Pichardo, B., Weilbacher, P., \& Mirabel, I.~F.\ 2000, \aj,
120, 1238

\bibitem[2005]{Elmegreen}
Elmegreen, B. G., 2005, to appear in Island Universes: Structure
and Evolution of Disk Galaxies, Springer, astro-ph/0510245

\bibitem[2002]{Garrido} Garrido, O., Marcelin,
M., Amram, P., \& Boulesteix, J.\ 2002, \aap, 387, 821

\bibitem[1996]{Heyl} Heyl, J.~S., Hernquist,
L., \& Spergel, D.~N. 1996, \apj, 463, 69

\bibitem[1995]{Hibbard} Hibbard, J.~E., \&
Mihos, J.~C., 1995, \aj, 110, 140

\bibitem[2004]{Jaeger}
J{\"a}ger, K., Ziegler, B.~L., B{\"o}hm, A., Heidt, J.,
M{\"o}llenhoff, C., Hopp, U., Mendez, R.~H., \& Wagner, S. 2004,
\aap, 422, 907

\bibitem[2004]{Kannappan} Kannappan, S.~J.,
\& Barton, E.~J. 2004, \aj, 127, 2694

\bibitem[2005]{Kapferer}
Kapferer, W., Knapp, A., Schindler, S., Kimeswenger, S., \& van
Kampen, E. 2005, A\&A 438, 87

\bibitem[2006]{Kapferer2}
Kapferer, W., Kronberger, T., Schindler, S., Ziegler, B.L., B\"ohm
A. 2006, A\&A, 446, 847

\bibitem[2006]{Karachentsev}
Karachentsev, I.D., Karachentseva, V.E., Huchtmeier, W.K. 2006,
A\&A, 451, 817

\bibitem[2003]{Mendes} Mendes de
Oliveira, C., Amram, P., Plana, H., \& Balkowski, C. 2003, \aj,
126, 2635

\bibitem[2006]{Metevier}
Metevier A.J., Koo D.C., Simard L., Phillips A.C. 2006,
astro-ph/0601671, accepted for publication in ApJ

\bibitem[1994]{Mihos}
Mihos, J.~C., \& Hernquist, L. 1994, APJ, 437, 611

\bibitem[2003]{Milvang} Milvang-Jensen,
B., Arag{\'o}n-Salamanca, A., Hau, G.~K.~T., J{\o}rgensen, I., \&
Hjorth, J. 2003, \mnras, 339, L1

\bibitem[1998]{Mo}
Mo, H.J., Mao, S., White, S.D.M. 1998, MNRAS 295, 319

\bibitem[2003]{Naab}Naab, T., \& Burkert,
A. 2003, \apj, 597, 893

\bibitem[2006]{Nakamura} Nakamura, O.,
Arag{\'o}n-Salamanca, A., Milvang-Jensen, B., Arimoto, N., Ikuta,
C., \& Bamford, S.~P. 2006, \mnras, 366, 144

\bibitem[1999]{Rubin}
Rubin, V.~C., Waterman, A.~H., \& Kenney, J.~D.~P. 1999, \aj, 118,
236

\bibitem[2001)]{Sofue}
Sofue, Y., \& Rubin, V.\ 2001, ARAß\&A, 39, 137

\bibitem[2003]{Springel}
Springel, V., Hernquist, L. 2003, MNRAS 333, 649

\bibitem[2005]{Springel}
Springel, V., Di Matteo, T., Hernquist, L. 2005, MNRAS 361, 776

\bibitem[2005]{Springel}
Springel, V. 2005, MNRAS 364, 1105

\bibitem[1999]{Swaters} Swaters, R.~A.,
Schoenmakers, R.~H.~M., Sancisi, R., \& van Albada, T.~S. 1999,
\mnras, 304, 330

\bibitem[1977]{Tully}
Tully, R.~B., \& Fisher, J.~R.\ 1977, A\&A, 54, 661

\bibitem[2005]{Verheijen}
Verheijen, M.~A.~W., Bershady, M.~A., Swaters, R.~A, Andersen,
D.~R., \& Westfall, K.~B. 2005, to appear in the proceedings of
'Island Universes - Structure and Evolution of Disk Galaxies',
astro-ph/0510360

\bibitem[2001]{Vogt}
Vogt, N. P. 2001, in ESO astrophysics symposia, Deep Fields, eds.
S. Cristiani, A. Renzini, \& R. E. Williams, (Springer), 112

\bibitem[2003]{Ziegler}
Ziegler, B.~L., B\"ohm A., J\"ager, K., Heidt, J., \&
M\"ollenhoff, C. 2003, APJ 598, L87

\end{thebibliography}
\end{document}